\documentclass{article}

\usepackage{microtype}
\usepackage{graphicx}
\usepackage{subfigure}
\usepackage{booktabs} %
\usepackage{tablefootnote} %
\usepackage{array}
\usepackage{tabularx}
\usepackage{threeparttable}
\usepackage{booktabs} %
\usepackage{amsmath} %

\newcolumntype{L}[1]{>{\raggedright\arraybackslash}p{#1}}
\newcolumntype{R}[1]{>{\raggedleft\arraybackslash}p{#1}}

\usepackage{hyperref}

\usepackage[accepted]{conf}

\makeatletter
\renewcommand{\Notice@String}{}
\makeatother

\mlsystitlerunning{SAKURAONE}

\begin{document}

\twocolumn[
\mlsystitle{SAKURAONE: An Open Ethernet–Based AI HPC System and Its Observed Workload Dynamics in a Single-Tenant LLM Development Environment}

\mlsyssetsymbol{equal}{*}

\begin{mlsysauthorlist}
\mlsysauthor{Fumikazu Konishi}{equal,sakura-research}
\mlsysauthor{Yuuki Tsubouchi}{equal,sakura-research}
\mlsysauthor{Hirofumi Tsuruta}{equal,sakura-research}
\end{mlsysauthorlist}

\mlsysaffiliation{sakura-research}{Research Center, SAKURA internet Inc., Japan}

\mlsyscorrespondingauthor{Fumikazu Konishi}{f-konishi@sakura.ad.jp}

\mlsyskeywords{Machine Learning, MLSys}

\vskip 0.3in
\begin{abstract}
SAKURAONE is a managed high performance computing (HPC) cluster developed and operated by the SAKURA Internet Research Center. It builds on the \emph{KOKARYOKU PHY} bare metal GPU platform and is optimized for advanced workloads, including large language model (LLM) training.  
In ISC 2025 TOP500, SAKURAONE is ranked \textbf{49th} by HPL and is the only top 100 system that uses a fully open networking stack—\textbf{800~GbE} with \textbf{SONiC}—demonstrating the scalability of vendor-neutral technology.  
Measured performance is 33.95~PFLOP/s (HPL~Rmax), 396.295~TFLOP/s (HPCG), and 339.86~PFLOP/s on HPL-MxP with FP8. The system consists of 100 nodes, each with eight NVIDIA H100 GPUs and a 2~PB all-flash Lustre file system, interconnected via a rail-optimized 800~GbE leaf–spine fabric with RoCEv2.  
Through exclusive use by a single research project, we observed the characteristics of development-related jobs. Consistent with previous HPC studies, small-scale jobs dominated in number, while a few large-scale jobs accounted for most GPU resource time. As the project progressed, resource use shifted from large-scale to mid-scale jobs, reflecting a transition from initial large-scale training to iterative refinement. These observations illustrate the real-world utilization dynamics of GPU clusters under unified project workloads.
\end{abstract}
]

\printAffiliationsAndNotice{\mlsysEqualContribution} %

\section{Introduction}
SAKURA internet Inc.\ has specialized in server and cloud infrastructure services since its founding in the early days of the Internet. 
As cloud computing has become a core component of modern societal infrastructure, its importance continues to grow alongside advances in artificial intelligence (AI) and data-intensive industries.

\begin{figure}[t]
\centering
\includegraphics[width=0.94\linewidth]{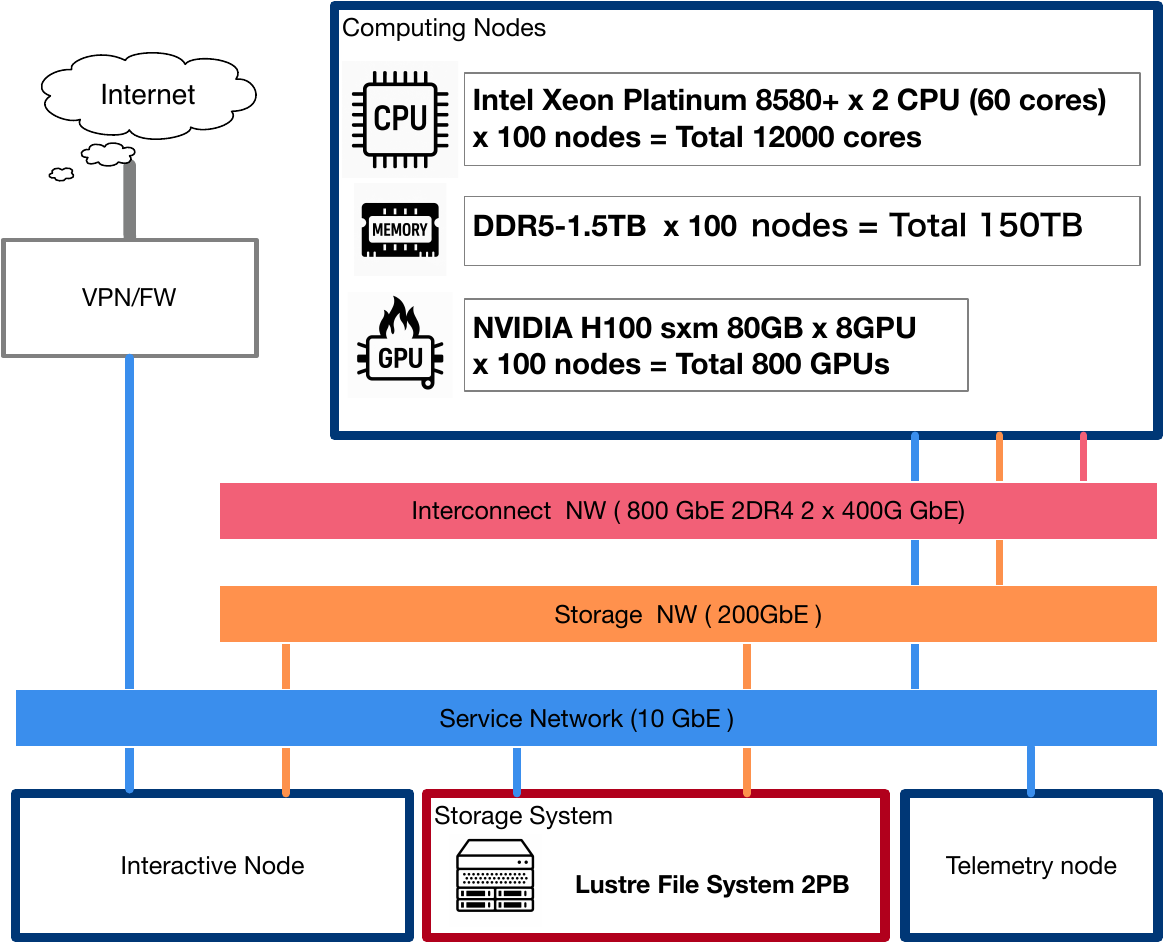}
\vskip -0.1in
\caption{SAKURAONE System Overview}
\label{fig:system_overview}
\vskip -0.1in
\end{figure}

To address emerging AI workloads, we present \emph{SAKURAONE}, a high performance computing (HPC) platform designed to meet the needs of industrial users in Japan. 
Figure~\ref{fig:system_overview} overviews the system: 100 compute nodes, each with eight NVIDIA H100 GPUs (800 GPUs total); a 2~PB all-flash Lustre storage subsystem for high-throughput, low-latency data access; a full-bisection-bandwidth interconnect in a rail-optimized topology over RoCEv2 for fast multi-node communication; and secure, high-speed VPN access to interactive front-end nodes for efficient remote use.

In this paper, we present an experience report from a production LLM development project encompassing continued pretraining, fine-tuning, and evaluation---rather than competing on maximum scale for one-shot pretraining.
All observations are drawn from a single-tenant, single-project setting.
While this limits claims of direct generalizability to multi-tenant production environments, the single-tenant setting reduces confounding factors such as cross-tenant contention and heterogeneous scheduling policies, enabling us to more clearly observe how workload characteristics and resource demands shift across development phases.
In addition to the tenancy aspect, the scale of our deployment warrants contextualization.
The majority of publicly available operational data today comes from hyperscale clusters with tens of thousands of GPUs.
In practice, however, many operators worldwide---including those in Japan---run mid-scale GPU clusters on the order of hundreds of GPUs as their primary platform for continued pretraining and fine-tuning.
Published data from such mid-scale production environments remain limited, especially for LLM-oriented workloads, and this paper helps fill that gap by reporting workload telemetry from an 800-GPU cluster.

\section{Background}
In the United States, major technology companies (``Big Tech'') have built inside-house AI infrastructures that enable sustained cutting-edge research and development. 
In contrast, while some Japanese companies are developing AI applications, these efforts are generally smaller in scale and often lack robust, dedicated computational backends.

Japan offers shared national and academic HPC resources—e.g., ABCI~3.0 operated by AIST \cite{takano2024abci30evolutionleading} and TSUBAME~4.0 at the Institute of Science Tokyo, which are available to industry \cite{10.1007/978-3-031-73716-9_16}. 
However, because such systems are shared with academic users, private-sector organizations can face challenges obtaining stable, predictable access at the capacity and time scales required for modern AI development. This creates a demand for comparably capable systems that are privately operated and dedicated to consistent commercial use.  
As part of efforts to improve the situation in which HPC resources are often constrained to closed vendor-specific system stacks, initiatives have emerged to pursue open, disaggregated architectures that increase transparency and flexibility.  

Since 2015, the push to open packet transport networks has accelerated. The adoption of the Switch Abstraction Interface (SAI) by the Open Compute Project has enabled clean hardware-software separation on switching platforms. This disaggregation improves flexibility and resilience, mitigates vendor lock-in, lowers barriers to entry, and lets hardware and software evolve independently; hyperscale data centers have widely embraced it along with pervasive storage and server virtualization. However, bringing virtualization comparable to optical transport demands a new architecture that decouples optical hardware from control software and digitizes the inherent analog characteristics of optical components.\cite{Nishizawa:20}

At the same time, open networking has matured. 
The Software for Open Networking in the Cloud (SONiC) network operating system (NOS) and the SAI standardize control of diverse switching ASICs, enabling a disaggregated, vendor-agnostic fabric across multiple hardware platforms. 
Beyond packet switching, SONiC-based NOS has also been demonstrated for open white-box optical transport equipment, including optical transponders, amplifiers, and protection switches, introducing an optical network line-card abstraction interface that provides a vendor-neutral unified line-card layer \cite{9979503}.

In hyperscale production (e.g. Microsoft Azure), the containerized architecture of SONiC and the FRRouting (FRR) control plane enable rapid evolution of features and broad validation of the community \cite{Yuan2018SONiC}. 
For AI/HPC fabrics, SONiC provides lossless Ethernet building blocks (Priority Flow Control (PFC) and Explicit Congestion Notification (ECN)) required by RoCEv2, and supports EVPN/VXLAN for scalable L2/L3 overlays \cite{Zhu2015DCQCN}\cite{Guo2016RDMAScale}. 
Together, these developments make open networking a credible option for production-grade AI interconnects.

\section{Motivation}
Japan’s leadership in AI research and industrial innovation is based on sustained access to large-scale computing. However, conventional HPC environments lack the elasticity, multi-tenancy, and observability required for modern AI workloads. To address this gap, our objectives are: (i) to advance national AI R\&D and industrial competitiveness through reliable large-scale compute access, and (ii) to reinforce next-generation HPC cloud infrastructure, including government cloud integration.

A key motivation in SAKURAONE is the use of open, disaggregated networking (e.g., SONiC on datacenter switches). This design improves vendor agility and supply chain resilience by decoupling NOS from the underlying ASICs, accelerates innovation through community-driven modular components, and enables lossless AI fabrics through RoCEv2 QoS (PFC, ECN). Scalable overlays (EVPN/VXLAN) further support multi-tenant AI workflows, while open management frameworks provide automation and observability essential for transparent large-scale operations.

These motivations yield concrete system requirements: predictable capacity for commercial users, scalable multi-GPU/multi-node training over lossless interconnects, sustained I/O for large datasets via all-flash Lustre, and secure, efficient remote access. The resulting architecture uses open networking to achieve balanced performance, flexibility, and cost efficiency throughout its lifecycle.

\section{Architecture}
\label{sec:architecture}
SAKURAONE comprises five subsystems that cover computation, interconnect, storage, secure access, and observability.
Interactive access and orchestration are handled by front-end nodes, while management and telemetry operate out-of-band in a read-only mode.
Training and storage networks are \emph{logically and physically separated} to minimize interference between collective traffic and I/O bursts.

\subsection{System Requirements}
To define the baseline capacity for SAKURAONE, we target \emph{continued pretraining} of a 70B-parameter LLM on approximately 300B tokens within four months.
Because practical development involves repeated runs and hyperparameter exploration, the system must sustain throughput sufficient for multiple overlapping full-size trainings without efficiency loss.

As a reference, the \emph{BLOOM-176B} model trained on the Jean Zay supercomputer used 384$\times$A100-80GB GPUs over 3.5 months (1.08M compute hours)~\cite{le_scao2022bloom}.
Using this as a baseline, we plan to shorten time-to-solution with \emph{Hopper-architecture} GPUs, whose tensor cores and Transformer Engine deliver roughly 2--3$\times$ higher per-GPU throughput for LLMs.
Under these assumptions, a cluster of about 100 compute nodes (8 GPUs/node, $\sim$800 GPUs total) provides sufficient headroom to meet the four-month target while enabling iterative and exploratory runs.

\subsection{Interconnect Requirements}
Scaling LLM training is fundamentally limited by inter-GPU communication.
In data- or hybrid-parallel regimes, each step synchronizes via collective operations, primarily Allreduce, imposing stringent bandwidth and latency constraints.
To sustain efficiency, the interconnect must support direct GPU-to-GPU data transfer, collective-optimized performance on Clos or rail-optimized fabrics, and co-design with software orchestration.

\subsubsection*{Direct GPU--to--GPU Data Movement}
GPUs must exchange data without host-memory staging or CPU mediation to reduce latency and overhead.
Peer-to-peer DMA between NICs and GPU memory (e.g., GPUDirect RDMA) is essential.
Within a node, high-bandwidth fabrics such as NVLink/NVSwitch should expose a unified topology that saturates inter-node links.

\subsubsection*{Collective-Oriented Performance (Clos \& Rail Topologies)}
Collectives dominate traffic; thus, networks require high bisection bandwidth, low tail latency, and stability under synchronized bursts.
Rail-optimized designs—multiple independent rails mapped to distinct channels—reduce congestion and improve throughput.
Libraries such as NCCL can stripe across rails using hierarchical algorithms.
Transport-level tuning (e.g., ECN control on RoCEv2) mitigates large collective incasts.

\subsubsection*{Performance Targets}
Communication must remain a minor fraction of step time; each GPU requires sufficient bandwidth and bounded collective latency.
Sustained efficiency should remain high at scale, parameterized by model and batch configuration.

\subsubsection*{Ethernet/RoCEv2 Engineering}
Ethernet fabrics can match HPC interconnects if engineered for lossless RDMA under collective bursts.
End-to-end ECN and DCQCN controls mitigate congestion without heavy PFC dependence, while QoS ensures background flows do not interfere with collectives.
High-speed links and multirail configurations are essential.

\subsubsection*{Scalability \& Topology-Aware Orchestration}
The interconnect must scale from rack to pod without redesign.
Hierarchical collectives should exploit physical locality, and scheduling should align placement with topology to confine communication to high-bandwidth domains.
Rail and domain awareness must be visible to runtimes and schedulers.

\subsection{Storage Requirements}
The storage subsystem must support concurrent large-scale training and data generation without interfering with the GPU fabric.
A dedicated storage network—physically and logically separate—prevents congestion spillover between checkpoint I/O and collectives.

A representative workload trains a 70B-parameter model on 32 nodes (256 GPUs), writing multi-terabyte checkpoints hourly.
Two to three such jobs may run concurrently, while up to 100 nodes perform data generation with sustained I/O.
These mixed patterns require a storage fabric with high aggregate bandwidth and strong tenant isolation.

A shared all-flash Lustre file system provides 2~PB usable capacity—reflecting 1~PB of expected output plus a 2$\times$ safety margin.
End-to-end throughput of $\sim$100~GB/s sustains simultaneous checkpointing and data generation without per-job degradation.
Each node connects via dual 400~GbE links to redundant storage switches for bandwidth and path diversity.
Metadata and object services are distributed for fault tolerance and availability.

The parallel file system absorbs concurrent I/O from all nodes, avoiding serialization bottlenecks and centralized chokepoints, ensuring balanced performance at scale.

\section{Implementation}
\label{sec:implementation}

\begin{figure*}[t]
\centering
\includegraphics[width=0.9\linewidth]{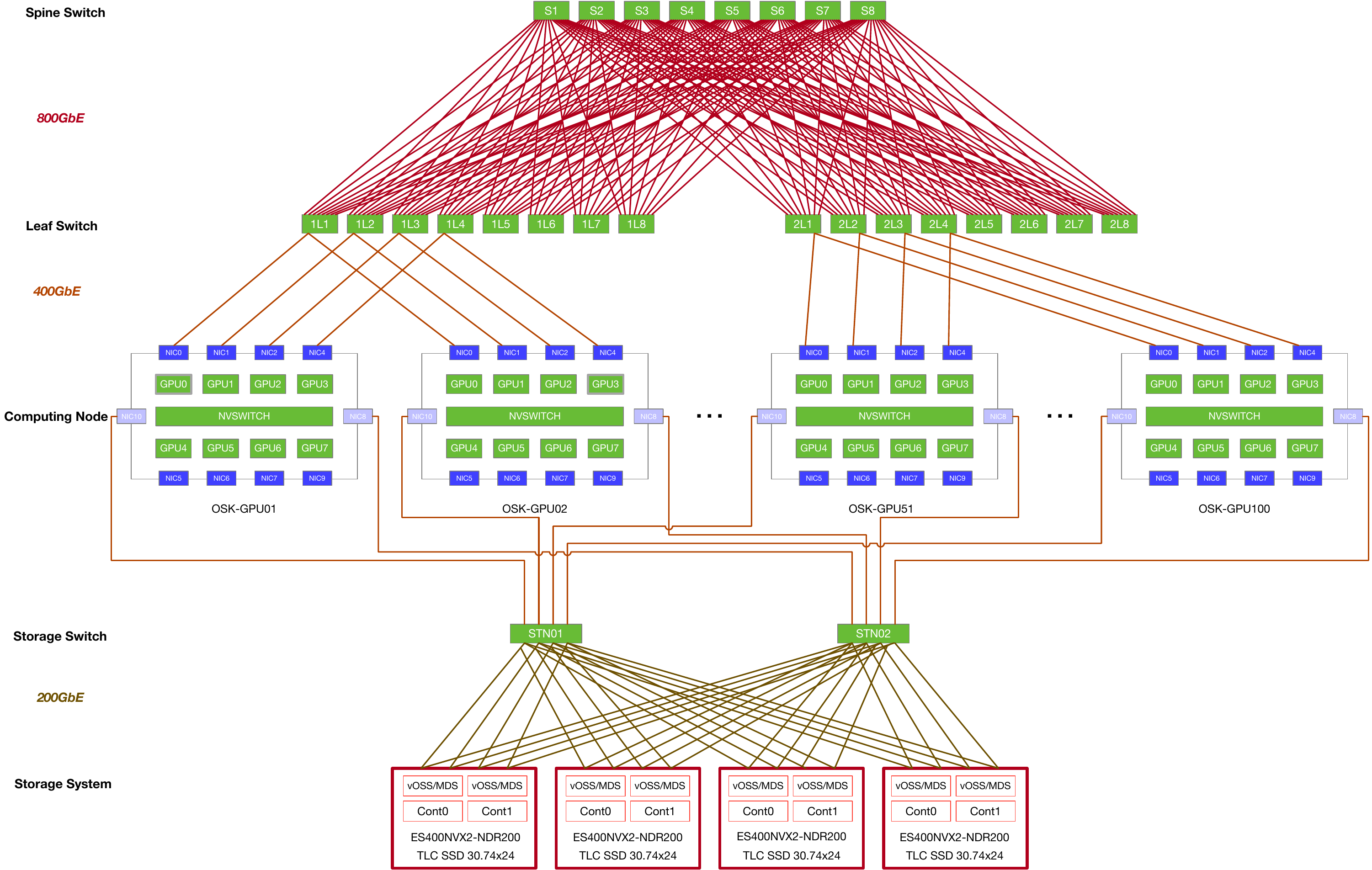}
\vskip -0.1in
\caption{SAKURAONE System Detail}
\label{fig:sakuraone}
\vskip -0.1in
\end{figure*}

\subsection{Compute Nodes}
Each compute node employs an 8U air-cooled GPU server equipped with two high-core-count CPUs, 1.5~TB of DDR5-5600 memory, and eight NVIDIA H100~SXM GPUs (80~GB each) interconnected via NVLink/NVSwitch, as summarized in Table~\ref{tab:compute_node}.
Local storage consists of four 7.68~TB NVMe SSDs for scratch space and two mirrored SAS drives for the system.
Networking provides eight 400~GbE ports for the GPU fabric, two 400~GbE ports for storage I/O, and a low-speed interface for management.

\begin{table}[t]
\caption{Compute Node}
\label{tab:compute_node}
\vskip 0.05in
\begin{center}
\begin{small}
\begin{tabular}{@{} l L{0.52\linewidth} @{}} %
\toprule
\textbf{Component} & \textbf{Specification} \\
\midrule
Chassis & Supermicro GPU SuperServer SYS-821GE-TNHR (8U, air-cooled) \\
CPU & \(2 \times\) Intel Xeon Platinum 8580+ (5th Gen, 60C/120T) \\
Memory & 1.5~TB DDR5-5600 \\
GPU & \(8 \times\) NVIDIA H100 SXM (80~GB), NVLink/NVSwitch \\
System storage & \(2 \times\) 372~GB SAS (mirrored) \\
Local scratch & \(4 \times\) 7.68~TB NVMe \\
Interconnect NICs & \(8 \times\) ConnectX-7, 400~GbE (GPU fabric) \\
Storage NICs & \(2 \times\) ConnectX-7, 400~GbE (I/O plane) \\
Management & 1~GbE (low-speed) \\
\bottomrule
\end{tabular}
\end{small}
\end{center}
\vskip -0.20in
\end{table}

\paragraph{GPU--NIC Affinity and Traffic Separation}
The proximity of PCIe / NVSwitch / NIC was profiled through \texttt{ nvidia-smi topo -mp}.
NIC0--NIC2, NIC4--NIC7, and NIC9 are \texttt{PIX}-connected to GPU0--GPU7 and reserved for inter-node collectives (NCCL/MPI over RoCEv2).
NIC8 and NIC10 (bonded as \texttt{mlx5\_bond\_0}) are dedicated to storage traffic.
NIC3 is reserved as a secondary network path, as classified in Table~\ref{tab:nic-usage}.

\begin{table}[t]
\caption{NIC Usage Classification and GPU Connectivity}
\label{tab:nic-usage}
\vskip 0.05in
\begin{center}
\begin{scriptsize}
\setlength{\tabcolsep}{1.5pt}
\begin{tabular}{@{} l l l l @{}}
\toprule
\textbf{NIC} & \textbf{Device} & \textbf{Primary Usage} & \textbf{GPU Connectivity} \\
\midrule
NIC0  & mlx5\_0      & Inter-node (RoCEv2)    & PIX (GPU0 PCIe domain) \\
NIC1  & mlx5\_1      & Inter-node (RoCEv2)    & PIX (GPU1 PCIe domain) \\
NIC2  & mlx5\_2      & Inter-node (RoCEv2)    & PIX (GPU2 PCIe domain) \\
NIC4  & mlx5\_4      & Inter-node (RoCEv2)    & PIX (GPU3 PCIe domain) \\
NIC5  & mlx5\_5      & Inter-node (RoCEv2)    & PIX (GPU4 PCIe domain) \\
NIC6  & mlx5\_6      & Inter-node (RoCEv2)    & PIX (GPU5 PCIe domain) \\
NIC7  & mlx5\_7      & Inter-node (RoCEv2)    & PIX (GPU6 PCIe domain) \\
NIC9  & mlx5\_11     & Inter-node/Management & PIX (GPU7 PCIe domain) \\
\midrule
NIC3  & mlx5\_3      & Secondary/Reserved       & NODE (GPU3 affinity) \\
NIC8  & mlx5\_8      & Storage (dedicated I/O)    & NODE (GPU7 affinity) \\
NIC10 & mlx5\_bond\_0 & Storage (bonded)           & Logical (multi-bridge) \\
\bottomrule
\end{tabular}
\end{scriptsize}
\end{center}
\vskip -0.20in
\end{table}

\subsection{Network Fabric}
Figure~\ref{fig:sakuraone} shows the rail--optimized leaf--spine fabric.
Each pod contains eight leaf switches; all leafs connect to all eight spine switches via 800~GbE inter-switch links.
Every compute node presents \(8 \times\) 400~GbE GPU-fabric links in the leaf set of the pod, resulting in uniform shortest-path connectivity between pods, as detailed in Table~\ref{tab:interconnect_network}.

\begin{table}[t]
\caption{Interconnect Network}
\label{tab:interconnect_network}
\vskip 0.05in
\begin{center}
\begin{small}
\begin{tabular}{@{} l L{0.52\linewidth} @{}} %
\toprule
\textbf{Item} & \textbf{Description} \\
\midrule
Network technology & Ethernet (GbE) \\
Port speed & 800~GbE (as \(2 \times\) 400~GbE) \\
Protocol & RoCEv2 (RDMA over Converged Ethernet) \\
Topology & Rail-optimized leaf--spine \\
Switch chassis & Edgecore AIS800-64O \\
Switch capability & 51.2~Tb/s full-duplex \\
Software stack & SONiC \\
Switch silicon & Broadcom Tomahawk~5 \\
Leaf switches & 16 chassis\footnotesize{~(two pods, eight leafs each)} \\
Spine switches & 8 chassis \\
\bottomrule
\end{tabular}
\end{small}
\end{center}
\vskip -0.20in
\end{table}

\subsection{Storage System}
Each node connects to the storage plane via two 400~GbE ports.
The shared Lustre tier is built on four DDN ES400NVX2-NDR200 servers, summarized in Table~\ref{tab:storage_system}.
Controllers provide active roles for object (OSS) and metadata (MDS) services; each server exposes eight 200~GbE interfaces redundantly wired to two storage switches for load balancing and failover.
A single-switch failure reduces aggregate bandwidth, but preserves service continuity.

\begin{table}[t]
\caption{Storage System}
\label{tab:storage_system}
\vskip 0.05in
\begin{center}
\begin{small}
\begin{tabular}{@{} l l r @{}} %
\toprule
\textbf{Component} & \textbf{Description} & \textbf{Count} \\
\midrule
Chassis    & DDN ES400NVX2 (all-flash) & 4 \\
Controllers & Active dual-controller & 2 \\
CPU        & Intel Ice Lake & 2 \\
NVMe bays  & 24 (PCIe Gen4) & 24 \\
Drive      & 30.72~TB TLC SSD & --- \\
Interfaces & 200~GbE per controller & 4 \\
\bottomrule
\end{tabular}
\end{small}
\end{center}
\vskip -0.20in
\end{table}

\subsection{System Software}
The SAKURAONE software stack is designed to (i) ensure production grade stability and reproducibility, (ii) support GPU-accelerated AI workloads, and (iii) enable portable, policy-compliant execution in a multiuser environment.

\begin{itemize}
  \item \textbf{Base OS and Runtime Environment.}
  The cluster runs Rocky~Linux~9.4 (RHEL-compatible) as a long-term supported base with regular security updates and wide HPC toolchain compatibility. A module-based environment provides multiple compiler, CUDA, and library versions for controlled upgrades and reproducible workflows.

  \item \textbf{Programming Models and GPU Libraries.}
  Standard parallel models (MPI, OpenMP) coexist with recent GPU toolchains. Multiple CUDA~12.x versions and optimized libraries (cuDNN, NCCL) support both traditional simulations and DNN training/inference.

  \item \textbf{Containers and Portability.}
  Singularity/Apptainer with Pyxis integration enables portable and dependency resolved execution in batch jobs. Users can run immutable container images consistently across nodes without elevated privileges.

  \item \textbf{Scheduling and Multi-Tenancy.}
  Slurm Workload Manager~22.05.9 manages resources with priority- and policy-based scheduling, reservations and job dependencies. Integration with monitoring tools provides real-time utilization insights and helps perform performance diagnostics under load.
\end{itemize}

Combining a stable OS, modular toolchains, containerized portability, and Slurm-based control, SAKURAONE delivers an HPC-grade environment optimized for both large-scale simulation and modern AI workloads with reproducible and policy-compliant execution.

\section{Evaluation}
We evaluated SAKURAONE using widely adopted HPC benchmarks to enable comparison with general-purpose clusters. Each benchmark targets different bottlenecks: compute throughput, memory and communication efficiency, mixed-precision AI performance, and storage I/O, providing a multifaceted view of system capability.

\subsection{Methodology and Setup}
Unless otherwise noted, runs were executed on SAKURAONE’s GPU partition with NVIDIA H100 (80~GB, SM~90) accelerators over a lossless RoCEv2 fabric. We report benchmark versions alongside problem sizes and process grids for reproducibility. Metrics follow community practice: sustained FLOP/s (Rmax) for HPL/HPL-MxP, validated GFLOP/s for HPCG, and IO500’s geometric-mean score with component bandwidth/IOPS.

\subsection{Dense Linear Algebra: HPL}
The High Performance Linpack (HPL) benchmark underpins the TOP500 list by solving dense linear systems in double precision and remains a standard indicator of sustained compute performance~\cite{Dongarra2001}. We used HPL-NVIDIA~25.4.0 with a matrix of size $N=2{,}706{,}432$, block size $NB=1024$, and a $16\times49$ process grid (784 GPUs), as summarized in Table~\ref{tab:hpl_summary}. SAKURAONE achieved \textbf{33.95~PFLOP/s} sustained (\textbf{43.31~TFLOP/s} per GPU). The single-GPU peak General Matrix Multiply (GEMM) rate was 55.34~TFLOP/s,
yielding a per-GPU efficiency of approximately 78.3\%.
The run completed in 389.23~s, indicating efficient scaling across the GPU fabric.

\begin{table}[t]
\caption{HPL Benchmark Summary}
\label{tab:hpl_summary}
\vskip 0.05in
\begin{center}
\begin{small}
\begin{tabular}{@{} L{0.50\linewidth} l @{}} %
\toprule
\textbf{Item} & \textbf{Value} \\
\midrule
HPL version & HPL-NVIDIA 25.4.0 \\
Matrix size $N$ & 2{,}706{,}432 \\
Block size $NB$ & 1024 \\
Process grid $(P\times Q)$ & $16\times49$ \\
Total GPUs / processes & 784 / 784 \\
Execution time & 389.23 s \\
Sustained performance (Rmax) & 33.95 PFLOP/s \\
Per-GPU performance & 43.31 TFLOP/s \\
Max single-GPU GEMM & 55.34 TFLOP/s \\
Per-GPU efficiency & 78.3\% \\
GPU SM count / peak clock & 132 / 1980 MHz \\
\bottomrule
\end{tabular}
\end{small}
\end{center}
\vskip -0.20in
\end{table}

\subsection{Memory/Communication-Limited Workloads: HPCG}
HPCG complements HPL by emphasizing sparse memory access and global communication~\cite{osti_1089988}. Using HPCG~3.1, we ran 784 distributed processes (16 threads each) on a global grid of $4096\times3584\times3808$, as detailed in Table~\ref{tab:hpcg_summary}. (\(\sim\)55.9~billion unknowns; \(\sim\)1.51~trillion nonzeros). Total memory footprint was 39.96~TB (35.17~TB for the linear system and CG). Observed peak memory bandwidth was 3.316~TB/s. The \emph{raw} rate reached 437{,}361~GFLOP/s (437.361~TFLOP/s); accounting for convergence overhead yielded 404{,}964~GFLOP/s (404.964~TFLOP/s). The \textbf{validated HPCG result} was \textbf{396{,}295~GFLOP/s} (\textbf{396.295~TFLOP/s}), indicating strong performance on communication- and memory-bound kernels.

\begin{table}[t]
\caption{HPCG Benchmark Summary}
\label{tab:hpcg_summary}
\vskip 0.05in
\begin{center}
\begin{small}
\begin{tabular}{@{} l L{0.38\linewidth} @{}} %
\toprule
\textbf{Item} & \textbf{Value} \\
\midrule
Benchmark version & HPCG 3.1 \\
Total processes / threads & 784 / 16 per process \\
Global problem $(n_x\times n_y\times n_z)$ & $4096\times3584\times3808$ \\
Unknowns / nonzeros & 55.9B / 1.51T \\
Total memory / system+CG & 39{,}961.4~GB / 35{,}169~GB \\
Peak memory bandwidth (obs.) & 3.316~TB/s \\
Raw compute rate & 437{,}361~GFLOP/s \\
With convergence overhead & 404{,}964~GFLOP/s \\
Validated HPCG result & 396{,}295~GFLOP/s \\
\bottomrule
\end{tabular}
\end{small}
\end{center}
\vskip -0.20in
\end{table}

\subsection{Mixed-Precision AI Linpack: HPL-MxP}
To reflect AI training characteristics, HPL-AI / HPL-MxP employs mixed-precision arithmetic with iterative refinement for accuracy~\cite{10.1109/SC.2018.00050}. Using HPL-MxP-NVIDIA~25.4.0, we solved with $N=2{,}989{,}056$, $NB=4096$, process grid $24\times32$ (768 GPUs), and Sloppy FP8 mode (type=1), as summarized in Table~\ref{tab:hpl_mxp_summary}. The observed \textbf{Rmax} was \textbf{339.86~PFLOP/s} (442.52~TFLOP/s per GPU). Isolating the LU factorization phase yielded \textbf{539.19~PFLOP/s} (702.07~TFLOP/s per GPU). Numerical validation passed with residual $5.01\times10^{-5}\ll1.6\times10^{1}$, confirming correctness under mixed precision.

\begin{table}[t]
\caption{HPL-MxP Benchmark Summary}
\label{tab:hpl_mxp_summary}
\vskip 0.05in
\begin{center}
\begin{small}
\begin{tabular}{@{} l L{0.38\linewidth} @{}} %
\toprule
\textbf{Item} & \textbf{Value} \\
\midrule
Benchmark version         & HPL-MxP-NVIDIA 25.4.0 \\
Matrix size $N$           & 2,989,056 \\
Block size $NB$           & 4096 \\
Process grid ($P \times Q$) & $24 \times 32$ \\
Total processes           & 768 \\
Peak clock frequency      & 1980~MHz \\
GPU SM version            & SM~90 \\
GPU SM count              & 132 \\
Observed $R_{\text{max}}$ & $3.3986 \times 10^{8}$~GFLOP/s \\
$R_{\text{max}}$ per GPU  & 442{,}520.81~GFLOP/s \\
LU-only                   & $5.3919 \times 10^{8}$~GFLOP/s \\
LU-only per GPU           & 702{,}074.99~GFLOP/s \\
Precision mode            & Sloppy FP8 (sloppy-type = 1) \\
Validation result         & PASSED ($5.01 \times 10^{-5} < 1.6 \times 10^{1}$) \\
\bottomrule
\end{tabular}
\end{small}
\end{center}
\vskip -0.20in
\end{table}

\subsection{Parallel Storage: IO500}
IO500 evaluates storage subsystems through bandwidth (IOR) and metadata (mdtest) kernels, reporting a geometric mean score and semiannual rankings~\cite{kunkel:2016:establishing}. We compare 10-node and 96-node results in Table~\ref{tab:io500_comparison}. While the 10-node setup delivered higher bandwidth in some IOR tests (e.g., \texttt{ior-easy-write/read}), the 96-node configuration excelled at metadata-heavy workloads (e.g., \texttt{mdtest-easy/hard-stat}, \texttt{find}), yielding a higher overall score. This suggests bandwidth saturation in the backend while metadata throughput scales with node count.

\begin{table}[htb]
\centering
\caption{IO500 Comparison: 10 Nodes vs.\ 96 Nodes}
\label{tab:io500_comparison}
\vskip 0.05in
\begin{center}
\begin{small}
\begin{tabular}{@{} L{0.3\linewidth} r r @{}} %
\toprule
\textbf{Benchmark} & \textbf{10 Nodes} & \textbf{96 Nodes} \\
\midrule
ior-easy-write (GiB/s)      & 262.91 (340.96 s) & 198.80 (355.68 s) \\
mdtest-easy-write (kIOPS)   & 204.44 (347.00 s) & 256.64 (363.81 s) \\
ior-hard-write (GiB/s)      & 15.84 (354.60 s)  & 24.61 (491.75 s)  \\
mdtest-hard-write (kIOPS)   & 120.84 (340.05 s) & 151.59 (332.84 s) \\
find (kIOPS)                & 1976.05 (56.39 s) & 2637.17 (54.01 s) \\
ior-easy-read (GiB/s)       & 365.71 (245.15 s) & 305.86 (231.13 s) \\
mdtest-easy-stat (kIOPS)    & 358.75 (197.40 s) & 463.13 (200.14 s) \\
ior-hard-read (GiB/s)       & 205.64 (31.23 s)  & 255.31 (48.82 s)  \\
mdtest-hard-stat (kIOPS)    & 262.43 (157.19 s) & 408.13 (124.54 s) \\
mdtest-easy-delete (kIOPS)  & 168.19 (422.12 s) & 198.91 (468.91 s) \\
mdtest-hard-read (kIOPS)    & 205.39 (200.53 s) & 310.87 (162.97 s) \\
mdtest-hard-delete (kIOPS)  & 92.29 (445.91 s)  & 111.28 (453.80 s) \\
\midrule
\textbf{Bandwidth score (GiB/s)} & \textbf{133.03} & \textbf{139.80} \\
\textbf{IOPS score (kIOPS)}      & \textbf{248.74} & \textbf{327.84} \\
\textbf{Total IO500 score}       & \textbf{181.91} & \textbf{214.09} \\
\bottomrule
\end{tabular}
\end{small}
\end{center}
\vskip -0.20in
\end{table}

\subsection{LLM Training: MLPerf}
MLPerf~\cite{mattson2020mlperf} is an industry-standard benchmark suite created by an open, nonprofit consortium of commercial and academic organizations to design a comprehensive benchmark for machine learning.
The MLPerf Training benchmark~\cite{mattson2020mlperftraining} measures the time required to train machine learning models to a defined quality target, providing an end-to-end performance metric that reflects real-world training scenarios.
To evaluate SAKURAONE's performance on contemporary LLM training workloads, we selected two benchmarks from the MLPerf Training v4.1 suite—GPT-3 175B pretraining and Llama 2 70B fine-tuning via Low-Rank Adaptation (LoRA)—and ran them informally on our system\footnotemark[1].
The parameters used in these benchmarks and their time-to-train measurements are presented in Tables~\ref{tab:mlperf_gpt3_summary} and~\ref{tab:mlperf_llama2_summary}, respectively.
Although these results are not official MLPerf submissions, they follow the benchmark specifications.

\footnotetext[1]{Unverified MLPerf\textsuperscript{\textregistered} Training v4.1 GPT-3 175B and Llama 2 70B LoRA fine-tuning benchmarks. Results not verified by MLCommons Association. The MLPerf name and logo are registered and unregistered trademarks of MLCommons Association in the United States and other countries. All rights reserved. Unauthorized use strictly prohibited. See \url{www.mlcommons.org} for more information.}

\begin{table}[t]
\caption{MLPerf Training (GPT-3) Benchmark Summary}
\label{tab:mlperf_gpt3_summary}
\vskip 0.05in
\begin{center}
\begin{small}
\setlength{\tabcolsep}{4pt}
\begin{tabular}{@{} l r r r @{}}
\toprule
\textbf{Item} & \textbf{32 N} & \textbf{64 N} & \textbf{96 N} \\
\midrule
Total GPUs           & 256 & 512 & 768 \\
Data Parallelism     & 4 & 8 & 6 \\
Tensor Parallelism   & 4 & 4 & 8 \\
Pipeline Parallelism & 16 & 16 & 16 \\
Virtual Pipelines & 6 & 6 & 6 \\
Global batch size    & 1024 & 1536 & 2304 \\
Micro batch size     & 2 & 2 & 6 \\
\midrule
Time-to-train (min)\textsuperscript{$\ast$}
                     & 105.31 & 58.30 & 41.86 \\
MFU (\%)             & 38.3 & 41.2 & 35.9 \\
Tokens/s/GPU         & 707.62 & 758 & 714.23 \\
TFLOPS/GPU           & 757.13 & 815 & 710.73 \\
\bottomrule
\multicolumn{4}{@{}l}{\textsuperscript{$\ast$}\scriptsize Unverified. 8 GPUs/node; CP\,=\,1, SP enabled for all configs.}
\end{tabular}
\end{small}
\end{center}
\vskip -0.20in
\end{table}

\begin{table}[t]
\centering
\caption{PyTorch Profiler breakdown for GPT-3 175B
(first pipeline stage, rank~0) at 32 and 64 nodes.}
\label{tab:profiler_breakdown}
\small
\begin{tabular}{@{}lrr@{}}
\toprule
\textbf{Metric} & \textbf{32\,N} & \textbf{64\,N} \\
\midrule
\multicolumn{3}{@{}l}{\textit{Temporal breakdown}\textsuperscript{$\ast$}} \\
\quad GPU compute & 81.7\% & 78.0\% \\
\quad Communication & 16.4\% & 19.3\% \\
\quad Idle & 1.9\% & 2.7\% \\
\quad Comm-comp overlap\textsuperscript{$\dagger$} & 72.3\% & 67.2\% \\
\midrule
\multicolumn{3}{@{}l}{\textit{NCCL kernel breakdown}\textsuperscript{$\ddagger$}} \\
\quad SendRecv (PP) & 91.2\% & 89.1\% \\
\quad ReduceScatter (TP) & 3.2\% & 3.5\% \\
\quad AllReduce (DP) & 3.8\% & 4.6\% \\
\quad AllGather (TP) & 1.8\% & 2.8\% \\
\bottomrule
\multicolumn{3}{@{}l}{\scriptsize\textsuperscript{$\ast$} Percentages sum to 100\% of step time.} \\
\multicolumn{3}{@{}l}{\scriptsize\textsuperscript{$\dagger$} Fraction of comm.\ time concurrent with compute.} \\
\multicolumn{3}{@{}l}{\scriptsize\textsuperscript{$\ddagger$} Shares relative to total NCCL kernel time.} \\
\end{tabular}

\end{table}

\paragraph{Communication Profiling.}
MFU is computed against the H100 SXM dense Tensor Core peak of
1{,}979~TFLOPS/GPU (without sparsity)~\cite{nvidia2023h100datasheet}.
We profiled the 32- and 64-node configurations using PyTorch Profiler
(Table~\ref{tab:profiler_breakdown}).
At 32 nodes, SendRecv (pipeline parallelism) dominates NCCL time
because PP\,=\,16 with VP\,=\,6 generates frequent inter-node
point-to-point traffic, while TP collectives stay on intra-node NVLink.
At 64 nodes, the same pattern holds but communication share
rises and overlap decreases, consistent with the cross-pod topology.

\paragraph{Scaling Across Configurations.}
The 64-node run keeps the same TP/PP/VP but doubles DP to 8 and
raises GBS to 1536 (Table~\ref{tab:mlperf_gpt3_summary}),
yielding MFU of 41.2\% (vs.\ 38.3\% at 32 nodes).
The allocation spans two pods connected via spine switches, and
Table~\ref{tab:profiler_breakdown} reflects this cross-pod penalty:
communication share and overlap both shift
(16.4\%$\to$19.3\% and 72.3\%$\to$67.2\%, respectively).
At 96 nodes, TP widens from 4 to 8 and DP changes to 6
(Table~\ref{tab:mlperf_gpt3_summary}), and MFU drops to 35.9\%;
profiling at this scale was not feasible within this work,
so a precise attribution is left to future work.

\begin{table}[t]
\caption{MLPerf Training (Llama 2 70B LoRA) Benchmark Summary}
\label{tab:mlperf_llama2_summary}
\vskip 0.05in
\begin{center}
\begin{tiny}
\begin{tabular}{@{} l l l l l @{}}
\toprule
\textbf{Item} & \textbf{1 Node} & \textbf{8 Nodes} & \textbf{64 Nodes} & \textbf{96 Nodes} \\
\midrule
Total GPUs                & 8    & 64   & 512  & 768 \\
Data Parallelism          & 2    & 8    & 64   & 96 \\
Tensor Parallelism        & 4    & 4    & 4    & 4 \\
Pipeline Parallelism      & 1    & 1    & 1    & 1 \\
Context Parallelism       & 1    & 2    & 2    & 2 \\
Sequence Parallelism      & True & True & True & True \\
Global batch size         & 8    & 8    & 64   & 96 \\
Micro batch size          & 1    & 1    & 1    & 1 \\
\midrule
Time-to-train (min)      & \shortstack[l]{28.44\\\normalfont (unverified)} & \shortstack[l]{4.79\\\normalfont (unverified)} & \shortstack[l]{1.94\\(\normalfont unverified)} & \shortstack[l]{1.26\\(\normalfont unverified)} \\
\bottomrule
\end{tabular}
\end{tiny}
\end{center}
\vskip -0.20in
\end{table}

\paragraph{Published-Reference Baseline.}
Table~\ref{tab:mlperf_eos_comparison} compares SAKURAONE with
NVIDIA Eos (DGX H100 SuperPOD, InfiniBand) using officially submitted
MLPerf Training v4.1 results~\cite{mlcommons2024trainingv41}.
The two systems differ in interconnect, node hardware, software stack,
and tuning, so ratios indicate relative positioning rather than
controlled equivalence.
At identical node counts, SAKURAONE's time-to-train is within
2--17\% of Eos (Table~\ref{tab:mlperf_eos_comparison}).
The 96-node Eos entry is a linear-scaling extrapolation
(favorable to Eos; see table note).

\begin{table}[t]
\caption{Published-Reference Comparison: SAKURAONE vs.\ NVIDIA Eos (MLPerf Training v4.1, Time-to-Train in Minutes)}
\label{tab:mlperf_eos_comparison}
\vskip 0.05in
\begin{center}
\begin{footnotesize}
\begin{tabular}{@{} l l r r r @{}}
\toprule
\textbf{Benchmark} & \textbf{Scale} & \textbf{Ours} & \textbf{Eos} & \textbf{Ratio} \\
\midrule
GPT-3 175B & 32 nodes & 105.31 & 96.66 & 1.09$\times$ \\
GPT-3 175B & 64 nodes & 58.30 & 49.80 & 1.17$\times$ \\
GPT-3 175B & 96 nodes & 41.86 & 33.20$^\dagger$ & 1.26$\times$ \\
\midrule
Llama 2 LoRA & 1 node & 28.44 & 27.93 & 1.02$\times$ \\
Llama 2 LoRA & 8 nodes & 4.79 & 4.57 & 1.05$\times$ \\
\bottomrule
\multicolumn{5}{@{}l}{\normalfont\scriptsize $^\dagger$ Linear extrapolation from Eos 64-node result ($49.80 \times 64/96$).} \\
\multicolumn{5}{@{}l}{\normalfont\scriptsize \phantom{$^\dagger$} Assumes perfect scaling (favorable to Eos).} \\
\end{tabular}
\end{footnotesize}
\end{center}
\vskip -0.20in

\end{table}

\subsection{Evaluation Summary}
Across dense, sparse/communication-limited, mixed-precision AI, storage, and LLM training workloads, SAKURAONE demonstrates balanced performance on its H100-based, lossless RoCEv2 fabric.
HPL sustains \textbf{33.95~PFLOP/s} on 784~GPUs, while HPCG validates \textbf{396.295~TFLOP/s}, indicating robust memory/collective behavior.
HPL-MxP reaches \textbf{339.86~PFLOP/s} overall and \textbf{539.19~PFLOP/s} in LU-only, highlighting mixed-precision throughput.
In storage, IO500 shows higher metadata scalability at 96 nodes and a higher overall score than the 10-node run, with bandwidth trends suggesting back-end saturation.
For the MLPerf Training benchmarks, GPT-3 175B pretraining achieved a time-to-train of \textbf{58.30~minutes (unverified)} on 64 nodes and \textbf{41.86~minutes (unverified)} on 96 nodes. The Llama 2 70B LoRA fine-tuning achieved \textbf{1.26~minutes (unverified)} on 96 nodes.
These results demonstrate performance comparable to official MLPerf Training v4.1 submissions at a similar scale, validating SAKURAONE's capability for production LLM workloads.
In particular, a published-reference comparison against NVIDIA Eos (DGX H100 SuperPOD, InfiniBand) shows that SAKURAONE achieves time-to-train within 2--17\% at identical node counts (Table~\ref{tab:mlperf_eos_comparison}).

\section{Observations}

\subsection{Scope and Timeline}
We conducted a Japanese medical-LLM project from \textbf{June 2024} to \textbf{March 2025}. Between \textbf{December 2024} and \textbf{March 2025}, we executed \emph{continued pretraining} (CPT) on \textbf{Llama-3.1-70B-instruct} and \textbf{Qwen2.5-72B-instruct} on SAKURAONE, followed by instruction tuning for EHR$\rightarrow$standard-code mapping. The present section summarizes operational observations from these runs.

\subsection{Key Observations}

\begin{figure}[t]
\centering
\includegraphics[width=\linewidth]{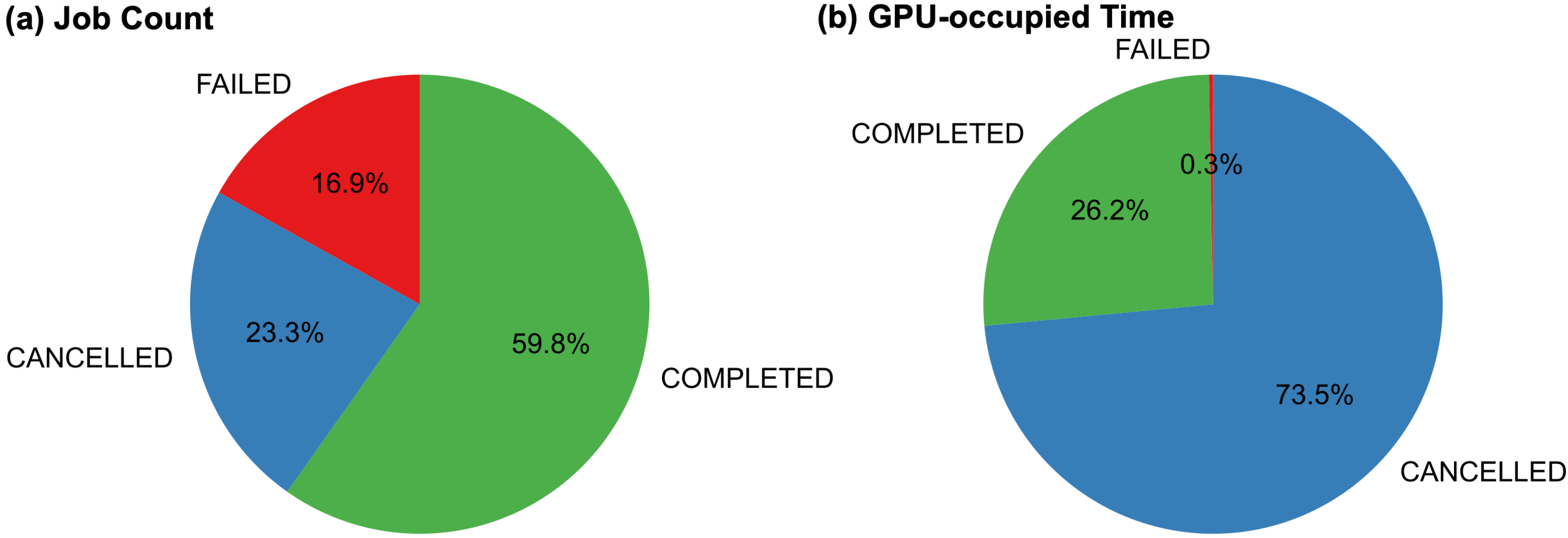}
\vskip -0.1in
\caption{Distribution of job states by (a) job count and (b) GPU-occupied time.}
\label{fig:state_distribution}
\end{figure}

\paragraph{Observation 1: User-initiated cancellations dominate GPU-occupied time, while failed jobs account for only 0.3\%.}
Figure~\ref{fig:state_distribution} shows the distribution of job states in terms of both job count and GPU-occupied time.
GPU-occupied time denotes the total duration during which jobs held GPU resources, regardless of their actual utilization.
It is computed as the product of the job’s runtime and the number of allocated GPUs; for example, a job running on four nodes (32 GPUs) for one hour corresponds to 32 GPU-occupied time.
This job state analysis reveals two key trends in cluster resource utilization.
First, CANCELLED jobs account for 73.5\% of the total GPU-occupied time.
In large-scale LLM training, it is often difficult to determine the optimal number of training steps in advance.
Practitioners therefore commonly set a conservatively high maximum step count and monitor training progress in real time using loss curves and validation metrics, terminating jobs once convergence is reached or training becomes unproductive.
For multi-day or multi-week experiments, such early termination serves as an important resource-efficiency practice.
The high cancellation proportion therefore reflects an operational pattern in which users proactively manage experiments and avoid continuing unproductive training runs.
Second, while 16.9\% of all jobs ended in the FAILED state, these failed jobs accounted for only 0.3\% of total GPU-occupied time.
This suggests that most failures occurred early in job execution, before significant GPU time was consumed.

\begin{figure}[t]
\centering
\includegraphics[width=\linewidth]{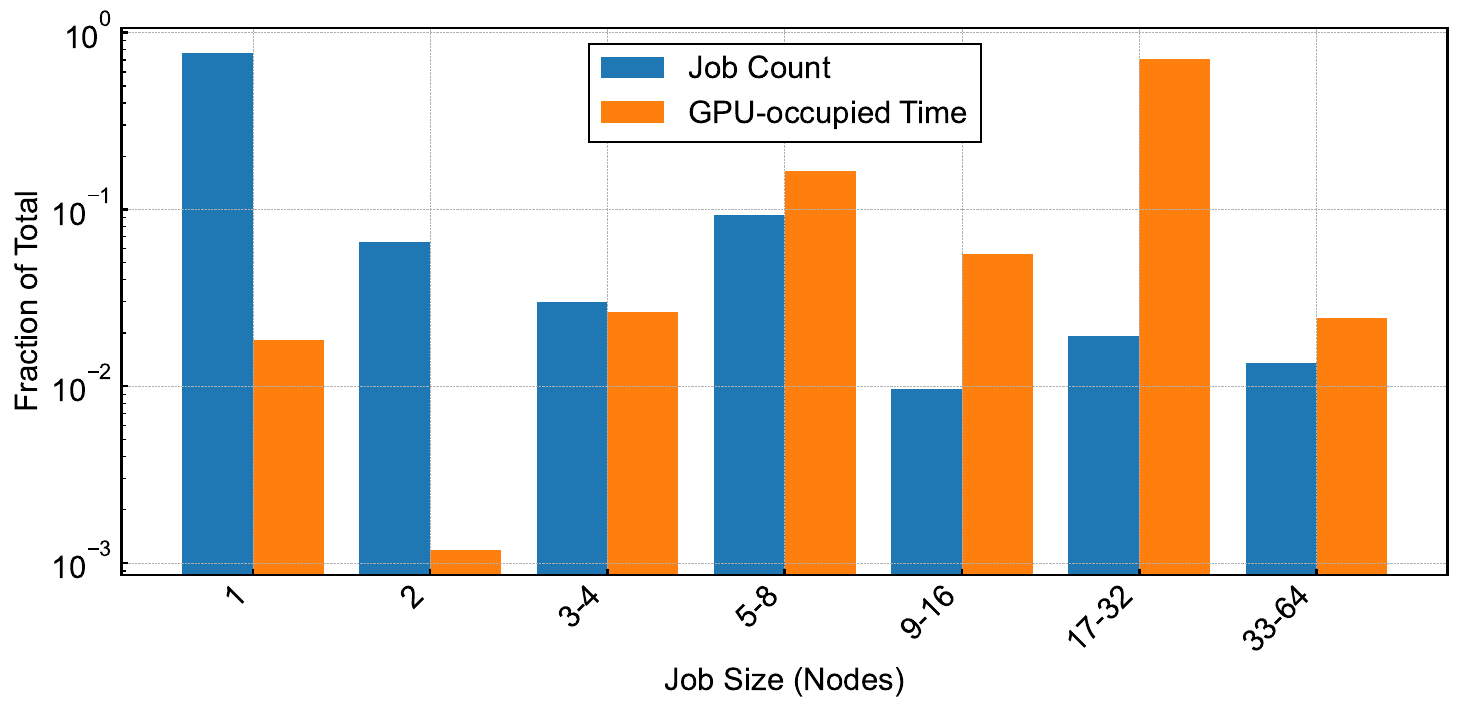}
\vskip -0.1in
\caption{Distribution of jobs by node count. Fraction of total job count (blue) and GPU-occupied time (orange) for each job size category.}
\label{fig:job_distribution}
\end{figure}

\paragraph{Observation 2: Small-scale jobs dominate in count, while large-scale jobs consume most of the GPU-occupied time.}
Figure~\ref{fig:job_distribution} shows the distribution of jobs by size (number of nodes), comparing job count and GPU-occupied time.
Among all jobs, 76.9\% ran on a single node and 86.4\% on four nodes or fewer. However, these small-scale jobs accounted for only 1.8\% and 4.6\% of total GPU-occupied time, respectively.
In contrast, jobs using 17 nodes or more represented only 3.3\% of job count but consumed 73.3\% of GPU-occupied time.
This pattern, in which small-scale jobs dominate numerically but large-scale jobs dominate resource consumption, has been reported in other large-scale AI infrastructure deployments~\cite{jeon2019analysis,kokolis2025revisiting}, suggesting it is a common characteristic of production GPU clusters.
This distribution reflects the dual role of GPU clusters, which must efficiently support both rapid experimental iteration (small-scale jobs) and production-scale training (large-scale jobs). Operationally, scheduling policies must balance responsiveness for numerous small-scale jobs against throughput for resource-intensive large ones.

\begin{figure}[t]
\centering
\includegraphics[width=\linewidth]{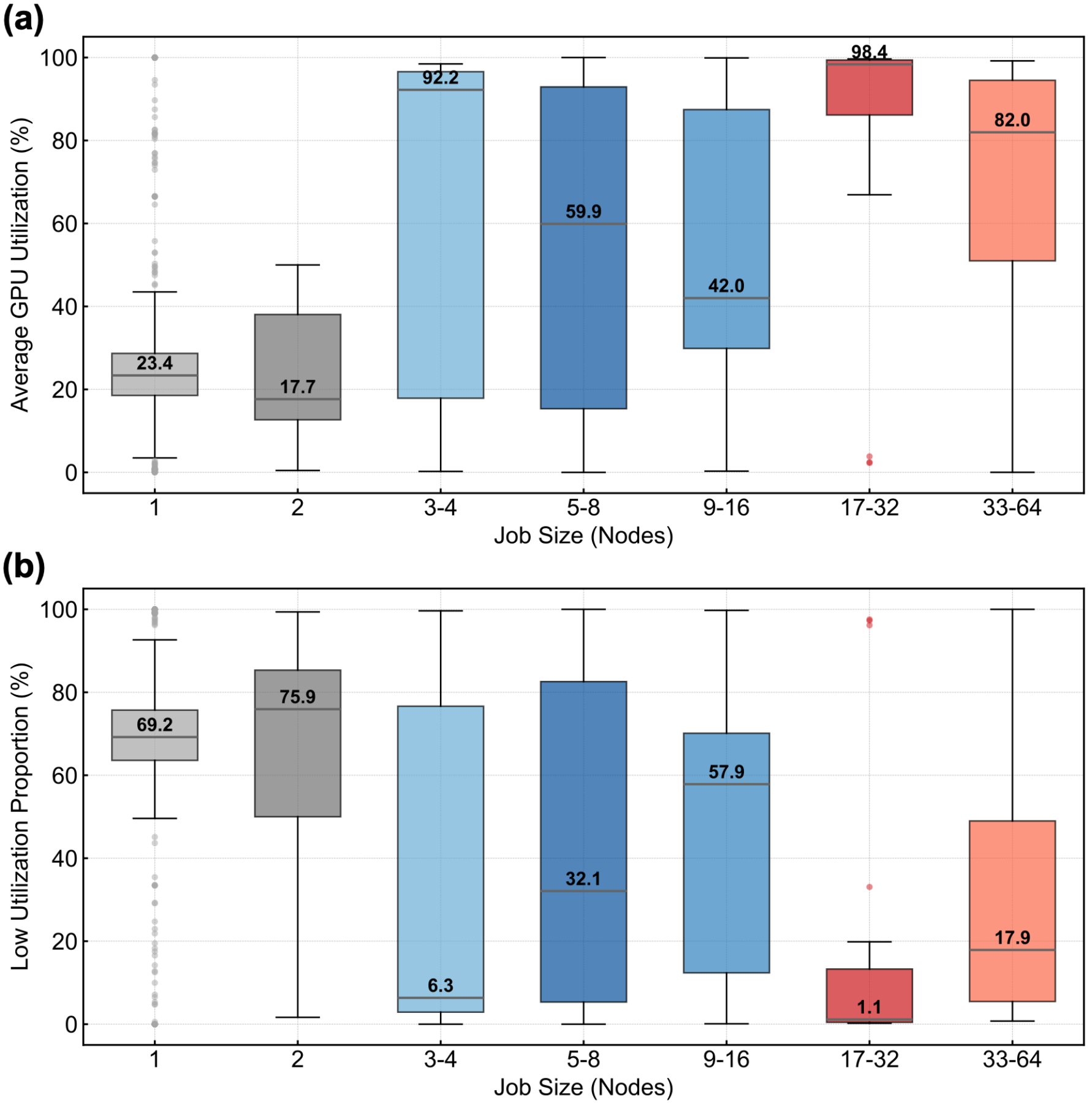}
\vskip -0.1in
\caption{Per-job GPU utilization by job size (nodes).
(a)~Distribution of average GPU utilization.
(b)~Distribution of the proportion of GPU-occupied time spent in low-utilization states (GPU utilization below 20\%).}
\label{fig:gpu_utilization}
\end{figure}

\paragraph{Observation 3: Large-scale jobs sustain high GPU utilization, whereas small-scale jobs spend the majority of their time in low-utilization states.}
To complement the GPU-occupied time analysis in Observation~2, we examine how effectively jobs actually utilize GPUs during their allocated time.
Following the methodology of Gao et al.~\cite{gao2024empirical}, we compute for each job an average GPU utilization across all allocated GPUs, and measure the proportion of GPU-occupied time spent in low-utilization states (below 20\%).
Figure~\ref{fig:gpu_utilization} reveals a clear distinction across job scales.
Large-scale jobs using 17--64 nodes, which primarily executed continued pretraining, achieved consistently high GPU utilization with minimal time in low-utilization states; in particular, jobs using 17--32 nodes achieved a median utilization of 98.4\% and spent only 1.1\% of their time in low-utilization states.
Mid-scale jobs using 3--16 nodes, associated with fine-tuning workloads among others, showed moderate utilization overall with notable variation across subcategories (median ranging from 42.0\% to 92.2\%).
Small-scale jobs (1--2 nodes) exhibited the lowest utilization, with medians of 23.4\% and 17.7\%, respectively, and spent 69.2\% and 75.9\% of their time in low-utilization states.
This pattern is consistent with the heterogeneous nature of small-scale tasks such as dataset preparation, model evaluation, and preprocessing, where CPU or I/O operations dominate over GPU computation.
These results provide quantitative evidence that GPU utilization varies systematically with job scale, reflecting the diverse computational demands across workload types within a single LLM development project.

\begin{figure}[t]
\centering
\includegraphics[width=0.7\linewidth]{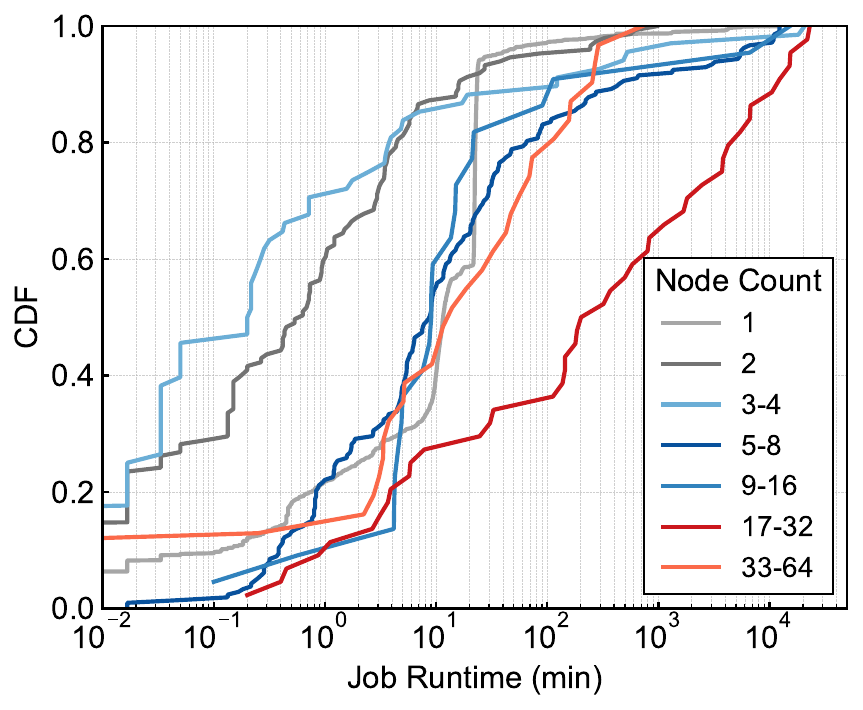}
\vskip -0.1in
\caption{Cumulative distribution of job runtimes by node count.}
\label{fig:job_runtime_cdf}
\end{figure}

\paragraph{Observation 4: Most jobs complete quickly, but large-scale jobs exhibit a long-tailed runtime distribution.}
Figure~\ref{fig:job_runtime_cdf} shows the cumulative distribution of job runtimes by node count.
The majority of jobs completed within a short duration, with runtimes typically within tens of minutes.
As job scale increased, however, the runtime distribution exhibited an increasingly long tail, with a growing fraction of long-running executions.
Notably, 13.6\% of jobs using 17-32 nodes exceeded one week of continuous runtime.
This long tail is primarily attributed to the CPT phase of the project, during which large-scale LLM training jobs were executed.

\begin{figure}[t]
\centering
\includegraphics[width=\linewidth]{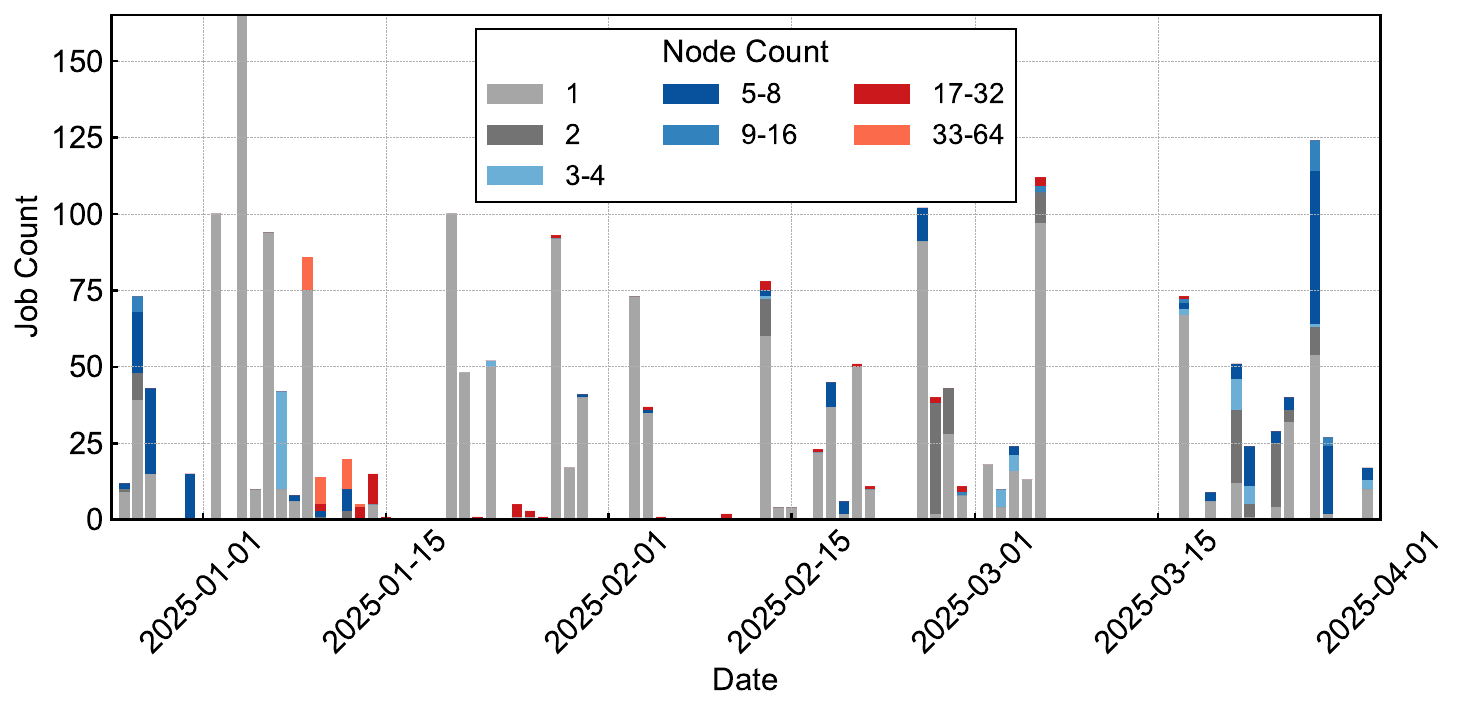}
\vskip -0.1in
\caption{Daily job submissions by node count.}
\label{fig:job_time_series_trend}
\end{figure}

\paragraph{Observation 5: Resource utilization shifts from large-scale to medium-scale jobs as the project progresses.}
Figure~\ref{fig:job_time_series_trend} shows the daily job submission counts grouped by node count throughout the project period.
Jobs are counted by their start date; long-running jobs spanning multiple days are counted only on the day they started.
From mid-January to early March 2025, large-scale jobs using 17-32 nodes (shown in red) were consistently submitted, reflecting the CPT phase in which various datasets and base models were trained.
Starting in mid-February 2025, however, medium-scale jobs using 3-16 nodes (shown in blue shades) gradually increased in frequency, likely marking a transition to the fine-tuning phase where the pretrained LLMs were adapted to downstream, task-specific datasets.
This temporal shift captures a typical development pattern within a single LLM project---an initial large-scale pretraining phase followed by a medium-scale fine-tuning phase.

\begin{table}[t]
\centering
\caption{Classification of 21 operational faults observed during the three-month period (January--March 2025), by component type.}
\label{tab:fault_classification}
\small
\begin{tabularx}{\columnwidth}{@{}Xrr@{}}
\toprule
\textbf{Fault Component} & \textbf{Count} & \textbf{Share (\%)} \\
\midrule
GPU (ECC / HW error / unresponsive) & 9 & 42.9 \\
NVLink / NVSwitch / PCIe switch & 4 & 19.0 \\
NIC / transceiver     & 1 &  4.8 \\
Interconnect switch (leaf/spine) & 5 & 23.8 \\
Storage switch        & 1 &  4.8 \\
Misconfiguration      & 1 &  4.8 \\
\midrule
\textbf{Total}        & \textbf{21} & \textbf{100.0} \\
\bottomrule
\end{tabularx}

\end{table}

\paragraph{Observation 6: GPU-related faults are the most frequent failure mode, but most faults are resolved by node-level restarts within minutes.}
During the three-month operational period, we observed 21 distinct fault events that disrupted or threatened normal cluster operation.
These events were collected from two independent incident channels (user-reported and platform-monitored), cross-referenced to remove duplicates, and verified by on-call engineers to exclude false positives.

Table~\ref{tab:fault_classification} classifies these faults by component type.
GPU-related faults were the most frequent category, accounting for 9 of 21 events (42.9\%), split between ECC memory errors (4 events) and other hardware errors or unresponsive GPUs (5 events).
Interconnect switch faults (leaf switch reboots, CPU failures, and MAC-learning anomalies---one of which manifested as cross-rail communication degradation) were the second most common at 5 events (23.8\%).
NVLink, NVSwitch, and PCIe switch faults within GPU servers together contributed 4 events (19.0\%).
The remaining 3 events included a NIC transceiver failure, a storage switch reboot, and an IP-address misconfiguration on a maintenance node.
While the majority of these faults originated in hardware components, the last event had a software or configuration root cause; we include it because it was operationally indistinguishable from hardware faults at the time of detection and required the same triage workflow.

Temporally, faults were concentrated in January 2025 (13 events), the first month of full-scale operation, and decreased in February (5 events) and March (3 events), suggesting an early burn-in period.
The dominant recovery method was node-level restart (warm or cold reboot), which resolved 10 of 21 events.
Three events required vendor-assisted hardware replacement (GPU tray, NVLink module, and NIC transceiver), with lead times ranging from days to weeks.
We do not report formal MTTF/MTTR statistics because our incident records are based on human-reported Slack logs with imprecise timestamps, which do not support reliable quantification of these metrics.

Note that the 0.3\% FAILED-job share (Observation~1) cannot be directly attributed to infrastructure faults alone, as the FAILED state also includes application-level errors, and some infrastructure faults led to manual cancellations rather than scheduler-reported failures.
At the 800-GPU scale, GPU memory and compute errors dominate the fault landscape, consistent with observations at larger scales~\cite{kokolis2025revisiting}.
The relatively short recovery times for the majority of faults---enabled by node-level isolation and Slurm's drain mechanism---indicate that modular server architectures and automated health checks can effectively contain the blast radius of individual component failures.

\begin{table}[t]
\centering
\caption{Interconnect bandwidth snapshot on the peak node at the NIC-peak instant for two representative jobs.
NIC peak is the single-port maximum 60-second full-duplex difference rate (intervals of 50--70\,s accepted) across eight inter-node 400\,GbE ports (nominal 100\,GB/s full-duplex each).
NVLink and PCIe values are per-GPU averages from the nearest DCGM sample ($\pm$2\,s).}
\label{tab:interconnect_bw}
\small
\begin{tabular}{@{}lrrrr@{}}
\toprule
 & \textbf{Nodes} & \textbf{NIC Peak} & \textbf{NVLink$^{\dagger}$} & \textbf{PCIe$^{\dagger}$} \\
 &                & \textbf{(GB/s)}   & \textbf{(GB/s)}             & \textbf{(GB/s)}            \\
\midrule
Job~A & 64 & 22.6 & 502.0 & 74.5 \\
Job~B & 32 & 18.9 & 114.5 & 17.4 \\
\bottomrule
\end{tabular}
\vskip 2pt
\raggedright\scriptsize $^{\dagger}$Per-GPU full-duplex average (TX+RX) over 8~GPUs on the peak node.

\end{table}
\vspace{-0.15in}

\paragraph{Observation 7: Per-port interconnect peaks reach 19--23\,GB/s, with inter-rail variation in some jobs.}
Table~\ref{tab:interconnect_bw} reports single-port peak bandwidth for two representative completed jobs, derived from NIC-side cumulative byte counters.
Job~A showed all eight inter-node ports uniformly at approximately 22.6\,GB/s, whereas Job~B exhibited per-rail asymmetry: six ports reached approximately 18.9\,GB/s while two remained near 8.0\,GB/s.
DCGM telemetry from the same instant confirms that intra-node interconnects were active in both cases: per-GPU NVLink throughput averaged 502.0\,GB/s (Job~A) and 114.5\,GB/s (Job~B), with PCIe at 74.5 and 17.4\,GB/s respectively (Table~\ref{tab:interconnect_bw}).
For Job~B, the PCIe asymmetry across GPUs mirrored the NIC-side rail imbalance---two GPUs recorded roughly 8.5\,GB/s versus approximately 20.3\,GB/s for the other six---suggesting that the non-uniformity was consistent across the interconnect stack at that instant.

Pod-level switch byte counters provide contextual corroboration but cannot be attributed to individual jobs, because switch telemetry aggregates all pod traffic.
At the Job~A peak instant, the pod-1 leaf switches recorded 45.2\,GB/s host-facing and 22.9\,GB/s spine-facing aggregate throughput; however, Job~A spanned both pods and only pod-1 counters were available, so this is a partial view.
For Job~B, pod-1 aggregate throughput fluctuated widely (host-facing 254.5--626.9\,GB/s; spine-facing 615.5--1231.7\,GB/s over a few seconds around the NIC peak), inflated by a concurrent 6-node job sharing the same pod.
The 60-second telemetry resolution smooths sub-second collective bursts: distributed DNN training traffic consists of repeated communication phases with markedly different network demand within each iteration~\cite{rajasekaran2024cassini,romero2022accelerating}, so our measurements may underestimate transient peak interconnect demand.
ECN marking rates and PFC pause counters were not collected during the study period, precluding direct attribution of observed bandwidth levels to congestion or its absence.

\section{Discussion}
The operational results of SAKURAONE offer several insights into both system design and real-world workload dynamics for large-scale LLM development.

\subsection{System Implications of an Open, Disaggregated Architecture}
SAKURAONE demonstrates that fully open Ethernet-based fabrics can deliver competitive performance for both HPC and AI workloads. Achieving 33.95~PFLOP/s on HPL and 339.86~PFLOP/s on HPL-MxP with a SONiC-managed 800~GbE fabric confirms that open networking can match the efficiency of proprietary interconnects. The RoCEv2-based multirail topology maintained lossless transport and observability through standard monitoring frameworks, validating the maturity of open networking for mission-critical AI systems.

However, achieving stability required precise coordination across firmware, kernel, and RDMA stack versions. Compared with InfiniBand-based clusters, tuning ECN thresholds, PFC behavior, and NCCL channel striping was crucial to prevent head-of-line blocking under synchronized bursts. These experiences suggest that, while open Ethernet fabrics are now viable, they still require deep cross-layer expertise for optimal collective performance.

\subsection{RoCE Congestion-Control Tuning}
ECN and PFC parameters for the RoCEv2 fabric were determined through vendor-validated testing on a simplified two-tier leaf–spine topology, in which ECN~min, ECN~max, and marking probability were swept under RingAllReduce and AlltoAll collective traffic patterns.
Table~\ref{tab:roce_tuning} lists the adopted production values.
PFC buffer parameters were left at vendor defaults, which are pre-calibrated to minimize packet drop; ECN thresholds were tuned to prevent the DCQCN~\cite{Zhu2015DCQCN} controller from entering 100\% mark-rate saturation prematurely.
These values were applied at the start of production and remained unchanged throughout the January--March 2025 observation period.

\begin{table}[h]
\centering
\caption{RoCEv2 congestion-control parameters adopted in SAKURAONE.}
\label{tab:roce_tuning}
\small
\begin{tabular}{ll}
\toprule
\textbf{Parameter} & \textbf{Value} \\
\midrule
ECN min / max            & 2~MB / 10~MB \\
ECN max marking probability & 1\% \\
PFC priority queue       & 3 (DSCP-based QoS) \\
PFC Xoff threshold       & 36,570,285~bytes \\
PFC Xon offset           & 18,432~bytes \\
PFC headroom             & 36~MB (shared, all ports) \\
Shared-buffer mode       & Dynamic (alpha~=~1, 66\%) \\
\bottomrule
\end{tabular}

\end{table}

The ECN max marking probability of 1\% may appear conservative, as it could be seen as relying more on PFC than on ECN-based rate reduction.
Although this configuration yielded the highest throughput in our validation tests, its suitability may vary with workload characteristics; we regard these parameters as a validated starting point and plan to adjust them as workload profiles evolve.

Two operational rules apply:
(1)~ECN min/max thresholds must be set in proportion to the available switch buffer capacity; under-provisioned values cause premature mark-rate saturation and unnecessary throughput loss.
(2)~The vendor-provided PFC buffer profile should be left at vendor defaults unless explicitly advised otherwise, as it is already calibrated for the hardware's buffer architecture.

\subsection{Workload Dynamics and Resource Efficiency}
The workload analysis revealed a dual structure common to GPU clusters: Small jobs dominated in count, while large-scale jobs consumed most of the GPU time. In SAKURAONE, 77\% of the jobs ran on a single node, but jobs that used more than 17 nodes occupied over 70\% of the total GPU time. This ``long-tail pattern'' mirrors trends in multi-tenant hyperscale clusters, indicating that even in a unified project, iterative experimentation coexists with large-scale production training.

A notable feature was the high proportion of user-initiated cancelations --- more than 70\% of the total GPU time. Rather than instability, this reflects adaptive control: practitioners monitored loss curves in real time and terminated runs early once convergence or saturation was observed. This proactive cancelation improved overall efficiency by avoiding long-term, unproductive jobs. This feedback-driven usage differs from traditional batch-oriented HPC patterns and calls for schedulers that better accommodate interactive AI workflows.

\subsection{Temporal Transition in LLM Development}
The shift from large-scale pretraining to medium-scale fine-tuning illustrates a typical LLM development lifecycle. During the CPT phase, long-running multinode jobs dominated; later, fine-tuning tasks executed over 3--16 nodes with shorter durations became prevalent. This evolution implies that static resource allocation is suboptimal; infrastructure should enable elastic reallocation and quick turnaround between phases. SAKURAONE’s Slurm-based control and network segmentation effectively supported this transition without contention between the compute and I/O planes.

\subsection{Scheduling Implications for Heavy-Tailed, Phase-Shifting Workloads}
The coexistence of numerous short jobs and long-running large jobs (Figure~\ref{fig:job_distribution}) creates a scheduling challenge: large jobs occupy most GPU resources for days or weeks (Figure~\ref{fig:job_runtime_cdf}), potentially limiting the resources available for short jobs.
To address this challenge, we suggest that checkpoint-based preemption~\cite{gu2019tiresias,mahajan2020themis}, which has been studied in GPU cluster schedulers, could be effective.
Specifically, a scheduler that uses checkpoint-completion events of long-running large jobs as safe interruption points---temporarily running pending short jobs and then resuming the large job from the checkpoint---could reduce short-job wait times without sacrificing multi-day training progress, even in single-tenant operation.
Moreover, the phase shift from large-scale continued pretraining to mid-scale fine-tuning (Figure~\ref{fig:job_time_series_trend}) further suggests that cluster resource configurations should not remain static throughout a project's lifecycle but be adjustable so the operational mode can be tuned as the workload mix changes.

\subsection{Positioning Relative to Other GPU Clusters}
Unlike national shared facilities such as ABCI or TSUBAME, SAKURAONE operated as a dedicated single-tenant system. This exclusivity eliminated queueing delays and enabled rapid iteration, but also exposed new challenges: utilization fluctuated between pre-training peaks and idle periods. Future systems could mitigate this by introducing controlled multi-tenancy, allowing lightweight inference or data-generation tasks to fill idle slots while maintaining isolation and compliance.

To clarify the generalizability of our results, we distinguish setting-specific preconditions from tenancy-independent findings.
\emph{Setting-specific preconditions} include the absence of cross-tenant contention, simplified priority and queue policies, and minimal queueing delays---all of which stem directly from the single-tenant arrangement and would not hold in shared environments.
\emph{Tenancy-independent findings} include the skewed resource-consumption pattern and the phase-driven shift from large-scale continued pretraining to mid-scale fine-tuning.
Aspects that arise only under multi-tenancy---such as queueing behavior, fairness policies, and reservation mechanisms---were not observed in our setting and remain an important area for future investigation.

Regarding the skewed resource-consumption pattern---small-scale jobs dominating in count while large-scale jobs consume most of the GPU time---the same trend has also been reported in other large-scale GPU cluster studies~\cite{jeon2019analysis,kokolis2025revisiting}; our mid-scale observations are consistent with this finding.
Publicly available operational studies of mid-scale (roughly 100--1{,}000 GPU) production clusters remain limited, especially for LLM-oriented training and development workloads; our observations therefore provide a useful reference point for operators in this underrepresented regime.

\subsection{Lessons for Future AI–HPC Co-Design}
This project highlights several principles for next-generation AI–HPC systems:
\begin{enumerate}
  \item \textbf{Open standards at scale.} SONiC-based Ethernet fabrics can rival proprietary interconnects when properly tuned.
  \item \textbf{Elastic scheduling over static allocation.} Workload phases vary drastically in scale; Flexible orchestration is more effective than fixed capacity.
  \item \textbf{Observability and user control.} Real-time telemetry enables human-in-the-loop optimization, turning cancellations into efficiency gains.
  \item \textbf{Fault containment through modularity.} Observation~6 shows that GPU-related faults dominate (42.9\%) and most are resolved by node-level restarts via Slurm's drain mechanism, validating modular server design. Hot-spare nodes further reduce downtime for events requiring hardware replacement. Automated pre-job GPU health checks~\cite{kokolis2025revisiting} are a natural next step.
\end{enumerate}

\subsection{Limitations and Future Work}
This study covers a single-tenant, single-project deployment; workload patterns may differ under multi-user contention. The single-tenant setting limits the generalizability of scheduling and queueing observations to multi-tenant environments. The nine-month track focuses on LLM training and fine-tuning, leaving multimodal and retrieval-augmented tasks for future study. The planned extensions include finer-grained telemetry of GPU utilization, I/O latency, and energy metrics to assess power-to-throughput efficiency. Integration with national research clouds is also under consideration.

\subsection{Broader Implications}
Ultimately, SAKURAONE shows that Japan’s commercial sector can bridge the gap between traditional HPC and modern AI infrastructure. As the only TOP500 system within the top 100 that uses a fully open Ethernet stack, it demonstrates that sovereign, cost-efficient compute infrastructure can coexist with cutting-edge AI performance. The insights from its architecture and workload dynamics contribute not only to HPC design but also to sustainable, transparent AI development practices.

\section{Related Work}

The evolution of RDMA and Ethernet-based interconnects has significantly shaped the design of large-scale HPC and AI systems. Early RDMA deployments focused on HPC and storage workloads, offering low latency and reduced CPU overhead. As the scale of the cluster increased, new challenges emerged in congestion control, reliability, and operational management. Zhu et~al.\ introduced DCQCN as an adaptive congestion control mechanism for large-scale RDMA environments~\cite{Zhu2015DCQCN}, while Guo et~al.\ demonstrated the practicality of RDMA over commodity Ethernet at scale~\cite{Guo2016RDMAScale}. More recently, Gangidi et~al.\ described Meta’s production deployment of RoCEv2 networks for distributed AI training, highlighting advances in topology, routing and transport tuning~\cite{10.1145/3651890.3672233}. Hoefler et~al.\ further surveyed hyperscale Ethernet+RDMA deployments, discussing lossless operation, fault containment, and scalability limits~\cite{hoefler2023datacenterethernetrdmaissues}. 

Parallel efforts in network topology have explored cost-effective high-bandwidth solutions optimized for AI and HPC workloads. The Dragonfly design remains a foundational high-radix topology for large systems~\cite{ISCA.2008.19,MM.2009.5}. Wang et~al.\ proposed \emph{Rail-only}, a rail-optimized network for trillion-parameter LLMs~\cite{Wang_2024}, while Hoefler et~al.\ introduced \emph{HammingMesh} to achieve predictable latency and flexible scheduling in deep learning clusters~\cite{10.5555/3571885.3571899}. Pichetti et~al.\ benchmarked Ethernet interconnects for HPC/AI workloads, showing competitive performance with appropriate tuning~\cite{10820803}.  

\textbf{Difference:} SAKURAONE builds on these studies by demonstrating a fully open SONiC/SAI-based Ethernet fabric that achieves closed efficiency of the InfiniBand class in production. Unlike previous work focused on simulations or proprietary systems, our deployment provides empirical evidence from continuous LLM training, linking network design principles with single-tenant real-world telemetry and operational stability.

\section{Conclusion}
SAKURAONE demonstrates that an open, Ethernet-based architecture can achieve HPC- and AI-grade scalability comparable to proprietary interconnects. Using SONiC-managed RoCEv2 fabrics and an all-flash Lustre storage system, the cluster sustained over 33.9~PFLOP/s on HPL and 339.8~PFLOP/s on HPL-MxP while supporting continuous large-scale LLM training. Operational telemetry revealed distinct workload dynamics: small jobs dominated in count, but large-scale training consumed most GPU time, gradually shifting toward mid-scale fine-tuning as model development matured. These results validate the feasibility of open, disaggregated infrastructures for production AI and provide design guidance for next-generation national and industrial compute platforms. Future work will extend telemetry to include energy and utilization metrics and explore federated operation across research and government AI/HPC platforms.

\section*{Acknowledgement}
We are deeply grateful to Takashi Inoue and Kiyohiro Kurosawa of the Cloud Service Department at SAKURA internet Inc. for their extensive support in designing the advanced network system and constructing the physical infrastructure.
We would also like to express our sincere gratitude to Tomohide Hattori of Prunus Solutions Inc. for his invaluable technical guidance and expertise in system integration.
Additionally, we thank Kota Kakiuchi and Shoetsu Sato of ELYZA, Inc. for providing critical insights from the perspective of LLM researchers and developers, which significantly contributed to validating our hypotheses during the data analysis.

This work was supported by the Cross-ministerial Strategic Innovation Promotion Program (SIP), ``Integrated Health Care System'' (Grant Number JPJ012425).

\nocite{langley00}

\bibliography{references}
\bibliographystyle{conf}

\end{document}